\documentclass[aps,showpacs,eqsecnum,twocolumn,superscriptaddress]{revtex4}

\usepackage{amsmath}
\usepackage{latexsym}
\usepackage{graphicx}
\usepackage{envmath}

\begin{document}


\title{BSSN equations in spherical coordinates without regularization: 
vacuum and non-vacuum spherically symmetric spacetimes}


\author{Pedro J. Montero}
\affiliation{Max-Planck-Institute f{\"u}r Astrophysik,
Karl-Schwarzschild-Str. 1, D-85748, Garching bei M{\"u}nchen, Germany}

\author{Isabel Cordero-Carri\'{o}n}
\affiliation{Max-Planck-Institute f{\"u}r Astrophysik,
Karl-Schwarzschild-Str. 1, D-85748, Garching bei M{\"u}nchen, Germany}

\date{\today}

\begin{abstract}
Brown $[$Phys. Rev. D {\bf 79}, 104029 (2009)$]$ has recently introduced a covariant formulation of the 
BSSN equations which is  well suited for curvilinear coordinate systems. This 
is particularly desirable as many astrophysical phenomena are symmetric with 
respect to the rotation axis or are such that curvilinear coordinates adapt 
better to their geometry. However, the singularities associated with such 
coordinate systems are known to lead to numerical instabilities unless special 
care is taken (e.g., regularization at the origin). Cordero-Carri\'{o}n will 
present a rigorous derivation of partially implicit Runge-Kutta methods in 
forthcoming papers, with the aim of treating numerically
the stiff source terms in wave-like equations that may appear as a result of 
the choice of the coordinate system. We have developed a numerical code solving
the BSSN equations in spherical symmetry and the general relativistic 
hydrodynamic equations written in flux-conservative form. A key feature of the 
code is that it uses a second-order partially implicit Runge-Kutta method to 
integrate the evolution equations. We perform and discuss a number of tests to 
assess the accuracy and expected convergence of the code, namely a pure gauge 
wave, the evolution of a single black hole, the evolution of a spherical 
relativistic star in equilibrium, and the gravitational collapse of a spherical
relativistic star leading to the formation of a black hole. We obtain stable 
evolutions of regular spacetimes without the need for any regularization 
algorithm at the origin.
\end{abstract}




\maketitle

\section{Introduction}
\label{sec:introduction}
The 3+1 formulation of Einstein equations originally proposed by 
Nakamura~\cite{Nakamura87} and subsequently modified by 
Shibata-Nakamura~\cite{Shibata95} and Baumgarte-Shapiro~\cite{Baumgarte99}, 
which is usually known as the BSSN formulation, has become the most widespread 
used formulation in the numerical relativity community. This is due to its 
stability properties, and to the developments associated with gauge conditions 
and the puncture method which have proved essential to perform accurate and 
long-term stable evolutions of spacetimes containing black holes 
(BHs)~\cite{Campanelli06,Baker06}.

The main drawback of the BSSN formulation in its original form resides in the 
fact that it is particularly tuned for Cartesian coordinates, since this 
involves dynamical fields which are not true tensors and assumes that the 
determinant of the conformal metric is equal to one. Brown~\cite{Brown09} 
addressed this issue and introduced a covariant formulation of the BSSN 
equations which is well suited for curvilinear coordinate systems. This is 
particularly desirable as many astrophysical phenomena are symmetric with 
respect to the rotation axis (e.g., accretion disks) or are such that spherical
coordinates adapt better to their geometry (e.g., gravitational collapse). 

However, the singularities associated with the curvilinear coordinate systems 
are a known source of numerical problems. For instance, one problem arises 
because of the presence of terms in the evolution equations that behave like 
$1/r$ near the origin $r=0$. Although on the analytical level the regularity of
the metric ensures that these terms cancel exactly, on the numerical level this
is not necessarily the case, and special care should be taken in order to avoid
numerical instabilities. A similar problem appears also near the axis of 
symmetry in axisymmetric systems if curvilinear coordinate systems are used.

Several methods have been proposed to handle the issue of regularity in 
curvilinear coordinates. One possible approach is to rely on a specific gauge 
choice (i.e., the {\it polar$/$areal} gauge)~\cite{Bardeen83,Choptuik91}, but 
it has the obvious limitation of restricting the gauge freedom which is one of 
the main ingredients for successful evolutions with the BSSN formulation. An 
alternative method is to apply a regularization procedure. One such
regularization methods, presented by \cite{Rinne05}, enforces both the 
appropriate parity regularity conditions and local flatness in order to achieve
the desired regularity of the evolution equations. Such method has the 
advantage that it allows a more generic gauge choice, and has been explored by 
\cite{Alcubierre05,Ruiz08,Alcubierre10} who have performed several numerical 
simulations of regular spacetimes in spherical and axial symmetry. In 
particular, in~\cite{Alcubierre10}, the authors applied a regularization 
algorithm to the BSSN equations in spherical symmetry. A disadvantage of such a
regularization algorithm is that it is not easy to implement numerically both 
conditions simultaneously, and it requires the introduction of auxiliary 
variables as well as finding their evolution equations. This is an obstacle if 
one wants to perform 3D simulations of regular spacetimes with spherical 
coordinates.

Therefore, one would ideally like to use a numerical scheme that is able to 
integrate in time a system of equations like the BSSN, in curvilinear 
coordinates (with or without symmetries), without the burden of regularization 
in order to achieve the desired stability and robustness. Implicit or partially
implicit methods are used to deal with systems of equations that require a 
special numerical treatment in order to achieve stable evolutions. The origin 
of the numerical instabilities may be diverse. Stiff source terms in the 
equations can lead to the development of numerical instabilities, and with some
choices of the coordinate system, source terms may introduce factors which can 
be numerically interpreted as stiff terms (e.g., $1/r$ factors due to spherical
coordinates close to $r=0$ even when regular data is evolved). Recently, 
partially implicit Runge-Kutta (PIRK) methods for wave-like equations in 
spherical coordinates have been successfully applied~\cite{CC12} to the 
hyperbolic part of Einstein equations in the Fully Constrained 
Formulation~\cite{FCF}.

The first steps through the rigorous derivation of the PIRK methods
will appear in~\cite{CC-Toro} and a detailed description of the methods and their
properties will be derived in a forthcoming paper~\cite{CC-PIRK}. Motivated by 
these results, we have developed a numerical code solving the BSSN equations in
spherical symmetry and the general relativistic hydrodynamics equations written
in flux-conservative form~\cite{Banyuls97}. The code uses a second-order PIRK 
method to integrate the evolution equations in time, and we do not apply any 
regularization scheme at the origin. This approach has the additional 
advantages that it imposes no restriction at all on the gauge choice (one can 
therefore use the moving puncture gauge) and no special care should be taking 
in the transition between a regular spacetime and that containing a singularity
as it happens in the gravitational collapse of a star to a BH.

The paper is organized as follows. The formulation of Einstein equations, 
including the implementation of the puncture approach and gauge conditions, 
along with the formulation of the general relativistic hydrodynamic equations 
is briefly presented in Sec.~II. Sec.~III gives a short description of the PIRK
method used, while Sec.~IV describes the numerical implementation. Sec.~V 
discusses numerical simulations of a pure gauge wave, the evolution of a single
BH, the evolution of spherical  relativistic stars in equilibrium, and the 
gravitational collapse of a spherical relativistic star leading to the 
formation of a BH. A summary of our conclusions is given in Sec.~VII. We use 
units in which $c=G=M_{\odot}=1$. Greek indices run from 0 to 3, Latin indices 
from 1 to 3, and we adopt the standard convention for the summation over 
repeated indices. 

\section{Basic equations}
\label{equations}
We next give a brief overview of the formulation for the system of Einstein and
hydrodynamic equations as it has been implemented in the code.

\subsection{BSSN equations in spherical symmetry}
\label{feqs}
A reformulation of the ADM system, the BSSN 
formulation~\cite{Nakamura87,Shibata95,Baumgarte99}, has been implemented to 
solve Einstein equations. In particular, we solve the BSSN equations in the 
special case of spherical symmetry. We refer to~\cite{Alcubierre10} for a 
detailed description of the equations.

Under this symmetry condition the spatial line element is written as
\begin{equation}
	dl^2 = e^{4\chi}[a(r,t)\,dr^{2} + r^{2}\,b(r,t)\,d\Omega^{2}],
\end{equation}
where $d\Omega^2$ is the solid angle element,
$d\Omega^2 = d\theta^{2} + \sin^2\theta d\varphi^{2}$, $a(r,t)$ and $b(r,t)$ 
are the metric functions, and $\chi$ is the conformal factor defined as
\begin{equation}
	\chi = \frac{1}{12}{\rm{ln}}(\gamma/\hat{\gamma}),
\end{equation}
where $\hat{\gamma}$ is the determinant of the conformal metric. The conformal 
metric relates to the physical one by 
\begin{equation}
	\hat{\gamma}_{ij} = e^{-4\chi} \gamma_{ij}.
\end{equation}

Initially, the determinant of the conformal metric fulfills the condition that 
it equals the determinant of the flat metric in spherical coordinates 
$\mathring{\gamma}_{ij}$ (i.e. 
$\hat{\gamma}(t=0) = \mathring{\gamma} = r^{4}\sin^{2}\theta$). Moreover, we 
follow the so called ``Lagrangian'' condition $\partial_{t}{\hat{\gamma}} = 0$ 
(i.e. choosing $\sigma=1$ in Eqs. (2.4), (2.6), (2.7), (2.16) and (2.17)). The 
evolution equation for the conformal factor takes the form
\begin{equation}
\label{chi}
	\partial_{t} \chi =\beta^{r} \partial_{r}\chi +
\sigma\hat{\nabla}_{m}\beta^{m}-\frac{1}{6}\alpha K,
\end{equation} 
$K$ being the trace of the extrinsic curvature, $\alpha$ the lapse function, 
and 
\begin{equation}
\label{divbeta}
	\hat{\nabla}_{m}\beta^{m} = \partial_{r} \beta^{r}
+ \beta^{r} \left( \frac{\partial_{r}(ab^2)}{2ab^2} + \frac{2}{r} \right)
\end{equation} 
the divergence of the shift vector $\beta^{i}$. The evolution equations for the
conformal metric components are:
\begin{equation}
\label{a}
	\partial_{t} a = \beta^{r} \partial_{r}a + 2a\partial_{r}\beta^{r}- 
\frac{2}{3}\sigma a\hat{\nabla}_{m}\beta^{m}-2\alpha a A_{a},
\end{equation} 
\begin{equation}
\label{b}
\partial_{t} b =\beta^{r} \partial_{r}b + 2b\frac{\beta^{r}}{r}- 
\frac{2}{3}\sigma b\hat{\nabla}_{m}\beta^{m}-2\alpha b A_{b},
\end{equation}
where $\hat{A}_{ij}$ is the traceless part of the conformal extrinsic 
curvature, and 
\begin{equation}
\label{A_a}
	A_{a} \equiv \hat{A}^{r}_{r}\;,\;\;\;\;\;\;
A_{b} \equiv \hat{A}^{\theta}_{\theta}. 
\end{equation}

Note that as $\hat{A}_{ij}$ is traceless $A_{a}+2A_{b}=0$. The evolution equation for $K$ is:
\begin{align}
\label{K}
	\partial_{t} K  & =  \beta^{r} \partial_{r}K - \nabla^{2}\alpha +
\alpha(A_{a}^{2} + 2A_{b}^{2} + \frac{1}{3}K^{2}) \nonumber \\
    & + 4\pi\alpha(E+S_{a}+2S_{b}),
\end{align}
with the matter source terms measured by the Eulerian observers given by 
\begin{eqnarray}
	E&=&n_{\mu}n_{\nu}T^{\mu\nu}, \nonumber \\
	j_{i}&=&-\gamma_{i\mu}n_{\nu}T^{\mu\nu}, \nonumber \\
	S_{ij}&=&\gamma_{i\mu}\gamma_{j\nu}T^{\mu\nu}, 
\label{ecmatsourc}
\end{eqnarray}
$T^{\mu\nu}$ being the stress-energy tensor for a perfect fluid, which is 
written as a function of the rest-mass density $\rho$, the specific enthalpy 
$h$, the pressure $P$ and the fluid 4-velocity $u^{\mu}$,
\begin{equation}
	T^{\mu\nu}= \rho h u{^\mu}u^{\nu} + P g^{\mu\nu},
\end{equation}
and
\begin{equation}
\label{S_a}
	S_{a} \equiv S^{r}_{r}\;,\;\;\;\;\;\;
S_{b} \equiv S^{\theta}_{\theta}. 
\end{equation}
The Laplacian of the lapse function with respect to the physical metric is 
given by 
%
%

\begin{align}
\label{nablaalpha}
	\nabla^{2} \alpha & = \frac{1}{\alpha
  e^{4\chi}} \bigg[\left.\partial^2_{r}\alpha -\partial_{r}\alpha\left(\frac{\partial_{r}a}{2a} -\frac{\partial_{r}b}{b}-2\partial_{r}\chi-\frac{2}{r}\right)\right].
\end{align}

Next, the evolution equation for the independent component of the traceless 
part of the conformal extrinsic curvature, $A_a$, is given by
\begin{align}
\label{Aa}
	\partial_{t} A_{a} & = \beta^{r}\partial_{r}A_{a} -
\left(\nabla^{r}\nabla_{r}\alpha - \frac{1}{3}\nabla^{2}\alpha\right)
+ \alpha\left(R^{r}_{r} - \frac{1}{3}R\right) \nonumber \\
	& + \alpha K A_{a} - 16\pi \alpha(S_a - S_b),
\end{align}
where $R^{r}_{r}$ is the mixed radial component of the Ricci tensor, $R$ its 
trace, and $\nabla^{r}\nabla_{r}\alpha$ is written as 
\begin{equation}
	\nabla^{r}\nabla_{r}\alpha = \frac{1}{\alpha e^{4\chi}}\left[ 
\partial^2_{r}\alpha - \partial_{r}\alpha\left(\frac{\partial_{r}a}{2a} 
+ 2\partial_{r}\chi\right)\right].
\end{equation}

Finally, the evolution equation for $\hat {\Delta}^{r}$, the radial component 
of the additional BSSN variables 
$\hat {\Delta}^{i} = \hat{\gamma}^{mn} \hat {\Delta}^{i}_{mn}$ with 
$\hat {\Delta}^{a}_{bc} = \hat{\Gamma}^{a}_{bc}-\mathring{\Gamma}^{a}_{bc}$, is
given by 
\begin{align}
\label{Deltar}
	\partial_{t}\hat {\Delta}^{r} & = \beta^{r}\partial_{r}\hat{\Delta}^{r} 
- \hat{\Delta}^{r}\partial_{r}\beta^{r} + \frac{1}{a}\partial^{2}_{r}\beta^{r} 
+ \frac{2}{b}\partial_{r}\left(\frac{\beta^r}{r}\right) \nonumber \\
 &+ \frac{\sigma}{3}\left(\frac{1}{a}\partial_{r}(\hat{\nabla}_m\beta^{m}) 
+ 2\hat{\Delta}^{r}\hat{\nabla}_m\beta^{m}\right) \nonumber \\
 &- \frac{2}{a}(A_{a}\partial_{r}\alpha + \alpha\partial_{r}A_{a}) \nonumber \\
 &+ 2\alpha\left(A_{a}\hat{\Delta}^{r} - \frac{2}{rb}(A_{a}-A_{b})\right)
\nonumber \\
 &+ \frac{\xi \alpha}{a} \left[\partial_{r}A_{a} - \frac{2}{3}\partial_{r}K 
+ 6A_{a}\partial_{r}\chi \right. \nonumber \\
 & \left. +(A_{a}-A_{b})\left(\frac{2}{r}+\frac{\partial_{r}b}{b}\right) 
- 8\pi j_{r} \right],
\end{align}
where we take $\xi=2$. 

Note that in the simulations shown in Sec.~V we have evolved the quantity 
$X \equiv e^{-2\chi}$ instead of the conformal factor $\chi$ (although similar
conclusions can be drawn if the conformal factor $\chi$ is used instead). We 
replace Eq.~(\ref{chi}) by the following evolution equation for $X$:
\begin{equation}
\label{X}
	\partial_{t} X = \beta^{r} \partial_{r}X 
- \frac{1}{3}X (\alpha K - \sigma \hat{\nabla}_{m}\beta^{m}).
\end{equation} 

In addition to the evolution equations there are constraint equations, the 
Hamiltonian and the momentum constraints, which are only used as diagnostics of
the accuracy of the numerical evolutions:
\begin{align}
\label{ham1}
  \mathcal{H} & \equiv R - (A^{2}_{a} + 2A^{2}_{b})
+ \frac{2}{3}K^{2}-16\pi E = 0, \\
\label{mom1}
  \mathcal{M}^r &\equiv \partial_{r}A_{a} 
- \frac{2}{3}\partial{r}K + 6A_{a}\partial_{r}\chi \nonumber\\
  &+(A_a -A_b)\left(\frac{2}{r} + \frac{\partial_r b}{b}\right) - 8\pi j_r = 0.
\end{align}
%

\subsubsection{Gauge choices}
In addition to the BSSN spacetime variables, there are two more variables left 
undetermined, the lapse, $\alpha$, and the shift vector, $\beta^{i}$. The code 
can handle arbitrary gauge conditions, however unless otherwise indicated, we 
use the so called {``non-advective 1+log''} condition~\cite{Bona97a} for the 
lapse, and a variation of the {``Gamma-driver''} condition for the shift vector \cite{Alcubierre02a,Alcubierre10}. 

The form of this slicing condition is expressed as
\begin{equation}
\label{1+log1}  
	\partial_{t}\alpha = -2 \alpha K. 
\end{equation}
For the radial component of the shift vector, we choose the Gamma-driver 
condition, which is written as 
\begin{align}
\label{shift1}  
	\partial_{t}B^{r} & = \frac{3}{4}\partial_{t}\hat{\Delta}^{r},\\
\label{shift2}  
 \partial_{t}\beta^{r} & = B^r,
\end{align}
where the auxiliary variable $B^r$ is introduced.

\subsection{Formulation of the hydrodynamic equations}    
The general relativistic hydrodynamic equations, expressed through the 
conservation equations for the stress-energy tensor $T^{\mu\nu}$ and the 
continuity equation are:
\begin{equation}
\label{hydro eqs}
	\nabla_\mu T^{\mu\nu} = 0\;,\;\;\;\;\;\;
\nabla_\mu \left(\rho u^{\mu}\right) = 0.
\end{equation}
Following \cite{Banyuls97}, the general relativistic hydrodynamic equations are
written in a conservative form in spherical coordinates. The following 
definitions for the hydrodynamic variables are used:
\begin{equation}
	v^{r}\equiv \frac{u^{r}}{\alpha u^{t}}+\frac{\beta^{r}}{\alpha},
\end{equation}
\begin{equation}
\label{def2}
	W\equiv \alpha u^{t},
\end{equation}
where $W$ is the Lorentz factor. By defining the vector of unknowns, 
${\bf{U}}$, as 
\begin{equation}
	{\bf{U}}=\sqrt{\gamma}(D,S_{r},\tau),
\end{equation}
where the conserved quantities are
\begin{align}
	D & = \rho W,\\
	S_r & = \rho h W^2v_r,\\
	\tau & = \rho h W^2 - P - D,
\end{align}
and fluxes, ${\bf{F}}^{r}$, as
\begin{align}
{\bf{F}}^{r} & =\sqrt{-g}\left[D(v^{r}
  -\beta^{r}/\alpha), \right. \nonumber \\
&  \left. S_{r}(v^{r}-\beta^{r}/\alpha)+P, \right. \nonumber \\
& \left. \tau  (v^{r}-\beta^{r}/\alpha)+P v^r\right],
\end{align}
the set of hydrodynamic equations (\ref{hydro eqs}) can be written in 
conservative form as 
\begin{equation}
\label{cylcon}
	\partial_{t}{\bf{U}}+\partial_{r}{\bf{F}}^{r}={\bf{S}},
\end{equation}
where ${\bf{S}}$ is the vector of sources given by 
\begin{align}
	{\bf{S}} &= \sqrt{-g}\left[0, 
T^{00}\left(\frac{1}{2}(\beta^{r})^{2}\partial_{r}\gamma_{rr}
- \alpha\partial_{r}\alpha \right) \right. \nonumber\\
 & \left. + T^{0r}\beta^r\partial_{r}\gamma_{rr}+T^{0}_{r}\partial_{r}\beta^{r}
  +\frac{1}{2}T^{rr}\partial_{r}\gamma_{rr}, \right. \nonumber\\ 
 & (T^{00}\beta^{r}+T^{0r})(\beta^{r}K_{rr} - \partial_{r}\alpha)+T^{rr}K_{rr}
  \bigg].
\end{align}
To close the system of equations, we choose the $\Gamma$-law equation of state 
given by 
\begin{equation}
\label{EOS1}
	P=\left(\Gamma -1\right)\rho\epsilon ,
\end{equation}
where $\epsilon$ is the specific internal energy.
 
\section{PIRK methods}
Let us consider the following system of PDEs,
\begin{System}
u_t = \mathcal{L}_1 (u, v), \\
v_t = \mathcal{L}_2 (u) + \mathcal{L}_3 (u, v),
\label{e:system}
\end{System}
$\mathcal{L}_1$, $\mathcal{L}_2$ and $\mathcal{L}_3$ being general non-linear 
differential operators. Let us denote by $L_1$, $L_2$ and $L_3$ their discrete 
operators, respectively. $L_1$ and $L_3$ will be treated in an explicit way, 
whereas the $L_2$ operator will be considered to contain the unstable terms 
and, therefore, treated partially implicitly.

We use a Runge-Kutta (RK) method to update in time the previous 
system~(\ref{e:system}). Each stage of the PIRK method consists of two steps: i) the variable $u$ is 
evolved explicitly; ii) the variable $v$ is evolved taking into account the 
updated value of $u$ for the evaluation of the $L_2$ operator. This strategy 
implies that the computational costs of the methods are comparable to those of 
the explicit ones. The resulting numerical schemes do not need any analytical 
or numerical inversion, but they are able to provide stable evolutions due to 
their partially implicit component.

For the numerical simulations shown in the paper, we use the second-order PIRK 
scheme, which follows as:
\begin{System}
	u^{(1)} = u^n + \Delta t \, L_1 (u^n, v^n),  \\
	v^{(1)} = v^n + \Delta t \left[\frac{1}{2} L_2(u^n) +
          \frac{1}{2} L_2(u^{(1)}) + L_3(u^n, v^n) \right],
\end{System}
\begin{System}
	u^{n+1}  = \frac{1}{2} \left[ u^n + u^{(1)} 
+ \Delta t \, L_1 (u^{(1)}, v^{(1)}) \right], \\
	v^{n+1}  =   v^n + \frac{\Delta t}{2} \left[ 
L_2(u^n) + L_2(u^{n+1}) \right.  \\ 
\left. \hspace{2.5cm} +  L_3(u^n, v^n) + L_3 (u^{(1)}, v^{(1)}) \right].
\end{System}
In the first stage, $u$ is evolved explicitly; the updated value $u^{(1)}$ is 
used in the evaluation of the $L_2$ operator for the computation of $v^{(1)}$.
Once all the values of the first stage are obtained, we proceed to the final
one. Again, $u$ is evolved explicitly (using the values of the variables of the
previous time-step and previous stage), and the updated value $u^{n+1}$ is used
in the evaluation of the $L_2$ operator for the computation of $v^{n+1}$.

This scheme is applied to the hydrodynamic and BSSN evolution equations. We 
include all the problematic terms appearing in the sources of the equations in 
the $L_2$ operator. Firstly, the hydrodynamic conserved quantities, the 
conformal metric components, $a$ and $b$, the conformal factor, $\chi$, or the 
quantity $X$ (function of the conformal factor), the lapse function, $\alpha$, 
and the radial component of the shift, $\beta^r$, are evolved explicitly (as 
$u$ is evolved in the previous PIRK scheme); secondly, the traceless part of 
the extrinsic curvature, $A_a$, and the trace of the extrinsic curvature, $K$, 
are evolved partially implicitly, using updated values of $\alpha$, $a$ and 
$b$; then, the quantity $\hat{\Delta}^{r}$ is evolved partially implicitly, 
using the updated values of $\alpha$, $a$, $b$, $\beta^r$, conformal factor, 
$A_a$ and $K$; finally, $B^r$ is evolved partially implicitly, using the 
updated values of $\hat{\Delta}^{r}$. Matter source terms are always included 
in the explicitly treated parts. In Appendix~\ref{appendix}, we give the exact 
form of the source terms included in each operator.

The PIRK methods will be further described in a forthcoming paper~\cite{CC-PIRK} and 
derived up to third-order in $\Delta t$ (time-step), in such a way that the number of 
stages is minimized. These methods are based on stability properties for both 
the explicit and implicit parts, recovering the optimal SSP explicit RK 
methods~\cite{GoShu98} when the $L_2$ operator is neglected, i.e., partially 
implicitly treated parts are not taken into account. 

\section{Implementation}
\label{sec:BCs}
\subsection{Numerics}
Derivatives in the spacetime evolution equations are calculated using a 
fourth-order centered finite difference approximation in a uniform grid except 
for the advection terms (terms formally like $\beta^{r}\partial_{r}u$), for 
which an upwind scheme is used. We also use fourth-order Kreiss-Oliger 
dissipation \cite{Kreiss} to avoid high frequency noise appearing near the 
outer boundary.

We use a second-order slope limiter reconstruction scheme (MC limiter) to 
obtain the left and right states of the primitive variables at each cell 
interface, and a HLLE approximate Riemann solver~\citep{Harten83,Einfeldt88}.

\subsection{Boundary Conditions}
The computational domain is defined as $0\leq r \leq L$, where $L$ refers to 
the location of the outer boundary. We used a cell-centered grid to avoid that 
the location of the puncture at the origin coincides with a grid point in 
simulations involving a BH. At the origin we impose the conditions derived from
the assumption of spherical symmetry. At the outer boundary we impose radiative
boundary conditions \cite{Alcubierre02a} for the spacetime variables expressed 
as
\begin{equation}
	\partial_{t} f = -v\partial_{r}f - \frac{v}{r}(f-f_{0}),
\end{equation}
where $f_{0}$ is the background solution of the field and $v$ is the wave 
speed.
 
\subsection{Atmosphere treatment}
An important ingredient in numerical simulations based on finite difference 
schemes to solve the hydrodynamic equations is the treatment of vacuum regions.
The standard approach is to add an atmosphere of very low density filling these
regions~\cite{Font02c}. We follow this approach and treat the atmosphere as a 
perfect fluid with a rest-mass density several orders of magnitude smaller than
that of the bulk matter. The hydrodynamic equations are solved in the 
atmosphere region as in the region of the bulk matter. If the 
rest-mass density $\rho$ or specific internal energy $\epsilon$ fall below the 
value set for the atmosphere, these values are reset to have the atmosphere 
value of the primitive variables.

\section{Numerical Results}
\label{sec:vacuum-tests}

\subsection{Pure gauge dynamics}
We first consider the propagation of a pure gauge pulse using the same initial 
parameters as in~\cite{Alcubierre10}. The main difference with respect 
to~\cite{Alcubierre10} is that we do not regularize the origin and rely only on
the PIRK scheme to achieve a stable numerical simulation. The initial data are 
given by
\begin{align}
	\chi & = 0,\\
	a&=b=1,\\
	A_{a}&=A_{b}=K=0,\\
	\hat{\Delta}^{r} &= 0, \\
	\alpha &= 1 +
\frac{\alpha_{0}r^{2}}{1+r^{2}}\left[e^{-(r-r_0)^{2}}+e^{-(r+r_0)^{2}} \right],
\end{align}
with $\alpha_{0}=0.01$ and $r_{0}=5$. We evolve these initial data with a grid 
resolution of $\Delta r = 0.1$ and $\Delta t = 0.5 \Delta r$. We use zero shift
and harmonic slicing,
\begin{equation}
	\partial_{t}\alpha = -\alpha^{2}K.
\end{equation}
%

\begin{figure}
\includegraphics[angle=0,width=8.cm]{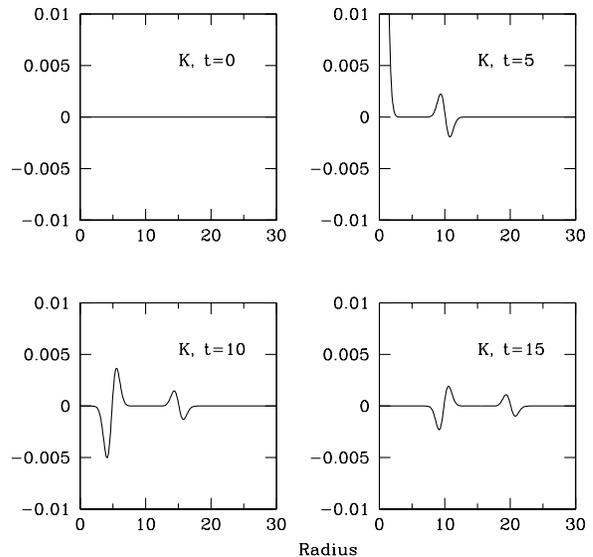} 
\caption{Trace of the extrinsic curvature, $K$, for a pure gauge pulse as a 
function of the radius at four different times.}
\label{fig1} 
\end{figure}

\begin{figure}
\includegraphics[angle=0,width=8.cm]{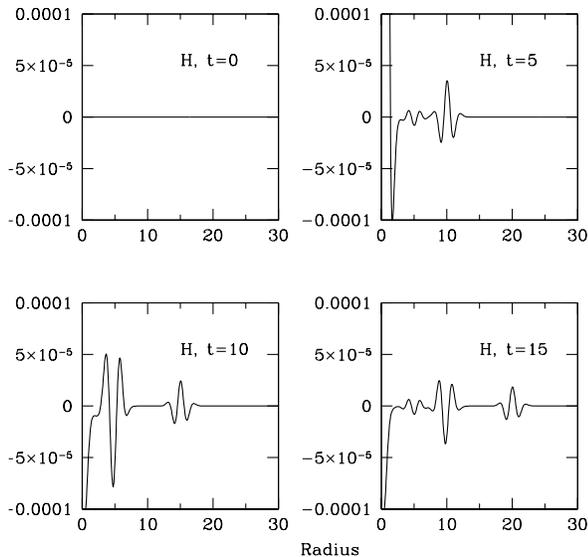} 
\caption{Hamiltonian constraint for a pure gauge pulse as a function of the 
radius at four different times.}
\label{fig2} 
\end{figure}

In Fig.~\ref{fig1}, we show the trace of the extrinsic curvature, $K$, as a 
function of the radius at four different times ($t = 0,5,10,15$). The initial 
pulse separates in two pulses propagating in opposite directions. The snapshots
of the evolution of the trace of the extrinsic curvature show that the 
evolution remains well behaved everywhere in the computational grid. We note 
that at $t=5$ the value of $K$ reaches a value of $~0.1$ at the origin, but 
later returns to zero when the pulse moves outwards as shown 
by~\cite{Alcubierre10}.

In Fig.~\ref{fig2}, we plot the Hamiltonian constraint (Eq.(\ref{ham1})) at 
four different times ($t = 0,5,10,15$). Although the largest violation of the 
Hamiltonian constraint occurs close to the origin (and is of the order of 
$~10^{-3}$ for the time frames shown in Fig.~\ref{fig2}), we find that it 
remains well behaved and there is no sign of any numerical instability despite 
the fact the initial data are regular at the origin and we do not impose any 
regularity conditions there.

\begin{figure}
\includegraphics[angle=0,width=8.cm]{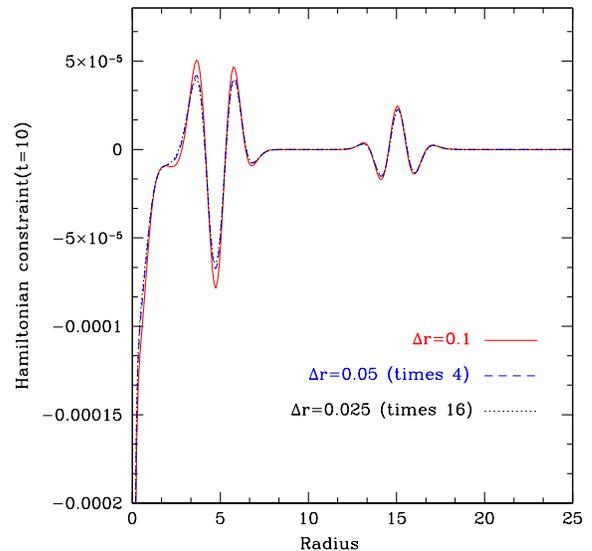} 
\caption{Hamiltonian constraint at $t=10$ for simulations of a pure gauge wave 
with three different resolutions $\Delta r = 0.1$, $\Delta r = 0.05$, and 
$\Delta r = 0.025$ rescaled by the factors corresponding to second-order 
convergence.}
\label{fig3} 
\end{figure}

In order to asses the convergence of the code, we have performed three 
simulations with resolutions $\Delta r =0.1$, $\Delta r =0.05$ and 
$\Delta r =0.025$. The Hamiltonian constraint violations rescaled by the 
factors corresponding to second order convergence at $t=10$ are plotted in 
Fig.~\ref{fig3}. All three lines overlap indicating that the code achieves the 
second-order convergence expected for the PIRK scheme used.

\subsection{Schwarzschild black hole}

The Schwarzschild metric in isotropic coordinates is used as initial data to 
test the ability of the code to evolve BH spacetimes within the moving puncture
approach. The initial data are such that the 3-metric is written as
\begin{equation}
\label{initdata}  
	dl^2=\psi^4(dr^2+r^{2}d\Omega^2), 
\end{equation}
where the conformal factor is $\psi=(1 + {M}/{2r})$, $M$ being the mass of the 
BH, which we set as $M=1$. Here $r$ is the isotropic radius. Initially the 
extrinsic curvature is $K_{ij}=0$.

\begin{figure}
\includegraphics[angle=0,width=8.cm]{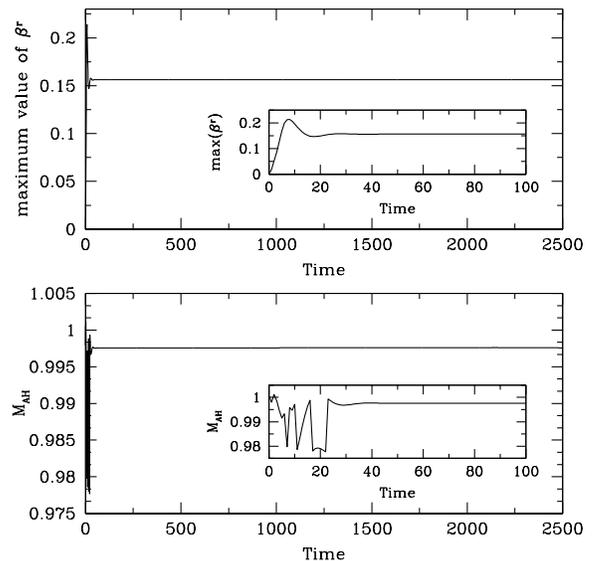} 
\caption{Time evolution of the maximum value of the radial shift $\beta^{r}$ 
(upper panel), and of the mass of the AH (lower panel) in the single puncture 
BH simulation.}
\label{fig4} 
\end{figure}

We evolve the single stationary puncture initial data with a precollapsed lapse
and initially vanishing shift vector. We use the gauge conditions given by 
Eqs.~(\ref{1+log1})--(\ref{shift2}), with a resolution $\Delta r =0.05$, 
$\Delta t= 0.5 \Delta r$ and $N_{r}=30000$ grid points to place the outer 
boundary sufficiently far way from the puncture so that errors from the 
boundary do not affect the evolution.

As pointed out by~\cite{Hannam07,Hannam07b}, the numerical slices of a 
Schwarzschild BH spacetime with these gauge conditions reach a stationary state
after $t\sim20$. This is shown in Fig.~\ref{fig4}, where the time evolution of 
the maximum value of the radial shift $\beta^{r}$ is displayed in the upper 
panel. After an initial phase in which the maximum value of the shift vector 
grows rapidly, it settles to a value of $\sim 0.15$ and we find almost no drift
until the end of the simulation at $t = 2500$. In the lower panel of 
Fig.~\ref{fig4}, we show the time evolution of the mass of the apparent horizon
(AH), defined as $M_{\rm {AH}}=\sqrt{{\mathcal{A}}/{16\pi}}$, where 
$\mathcal{A}$ is the area of the AH. We notice that $M_{\rm {AH}}$ is conserved
well during the evolution and the error at $t=2500$ is less than $0.2\%$

\begin{figure}
\includegraphics[angle=0,width=8.cm]{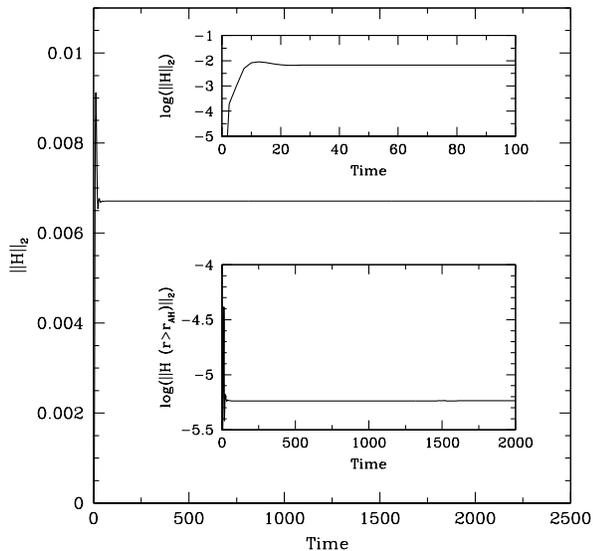} 
\caption{L2-norm of the Hamiltonian constraint in the single puncture BH 
simulation. The insets show the L2-norm during the initial phase, and the 
L2-norm computed outside the AH respectively.}
\label{fig5} 
\end{figure}

The Hamiltonian constraint violation results are displayed in Fig.~\ref{fig5}, 
which shows the L2-norm of the Hamiltonian constraint as a function of time. 
Both the initial phase driven by the gauge dynamics (see upper inset in 
Fig.~\ref{fig5}) and the stationary phase are clearly visible. The lower inset 
in Fig.~\ref{fig5} shows that the L2-norm computed outside the AH is about 
three orders of magnitude smaller than the L2-norm computed in the whole grid, 
which is to be expected as the largest spatial violation of the constraint 
occurs near the puncture (due to the finite differencing of the irregular 
solution).

\subsection{Spherical relativistic stars}

\begin{figure}
\includegraphics[angle=0,width=8.cm]{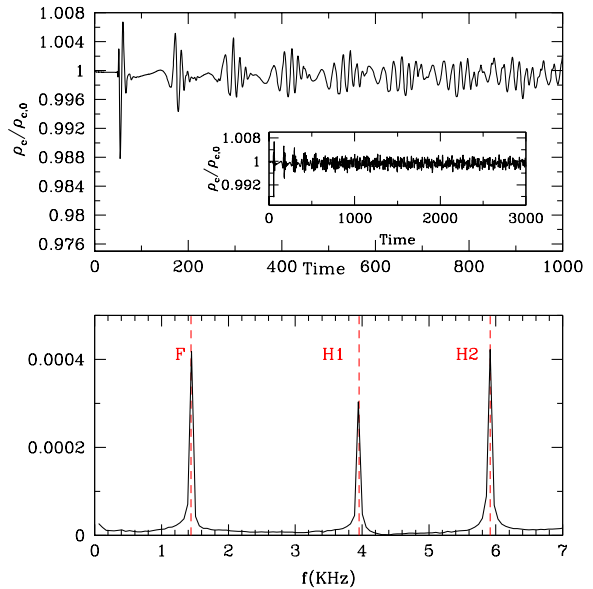}
\caption{Upper panel shows the time evolution of the normalized central density 
for an $M=1.4$, $\kappa=100$, $N=1$ polytrope. Power spectrum of the evolution 
of the central rest-mass density is shown in the lower panel. F, H1 and H2 
represent the frequency of the fundamental mode and the first two overtones 
computed by \cite{Font02c}.} 
\label{fig6} 
\end{figure}

For our first numerical simulation of the coupling of Einstein equations and 
the general relativistic hydrodynamic equations, we use the 
Tolman-Oppenheimer-Volkoff (TOV) solution. We focus on an initial TOV model 
that has been extensively investigated numerically by \cite{Font00c,Font02c}. 
This model is a relativistic star with polytropic index $N=1$, polytropic 
constant $\kappa=100$ and central rest-mass density 
$\rho_{c}=1.28\times 10^{-3}$, so that its gravitational mass is $M=1.4$, its 
baryon rest-mass $M_{*}=1.5$ and its radius $R=9.59$. 

We evolve these initial data with our non-linear code until $t=3000$ 
($\sim$17 ms). In the upper panel of Fig.~\ref{fig6} we plot the time evolution
of the central rest-mass density for a simulation with $\Delta r=0.025$ and 
$N_{r}= 4000$ until $t=1000$. In the inset we show the same quantity for the 
whole evolution. We observe that the truncation errors at this resolution are 
enough to excite small periodic radial oscillations, visible in this plot as 
periodic variations of the central density. We see that the damping of the 
periodic oscillations of the central rest-mass density is very small during the
whole evolution, which highlights the low numerical viscosity of the 
implemented scheme. 

By computing the Fourier transform of the time evolution of the central 
rest-mass density we obtain the power spectrum, which is shown with a solid 
line in the lower panel of Fig.~\ref{fig6}, while the dashed vertical lines 
indicate the fundamental frequency and the first two overtones computed by 
\cite{Font02c}. Note that the locations of the frequency peaks for the 
fundamental mode and the two overtones are in very good agreement, the 
relative error in the fundamental frequencies being less than 0.1\%.

The result of this simulation shows the ability of the scheme to maintain the 
numerical stability in long-term non-vacuum regular spacetime simulations in 
spherical coordinates without the need of an additional regularization at the 
origin.  
 
\begin{figure}
\includegraphics[angle=0,width=8.cm]{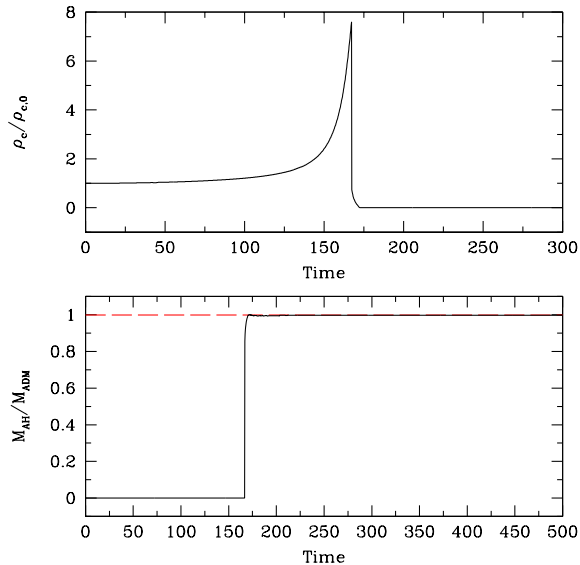} 
\caption{Time evolution of the normalized central density (upper panel) and mass 
of the AH in units of the ADM mass of the system (lower panel) for the 
collapse of a marginally stable spherical star to a BH.}
\label{fig7} 
\end{figure}

\begin{figure}
\includegraphics[angle=0,width=8.cm]{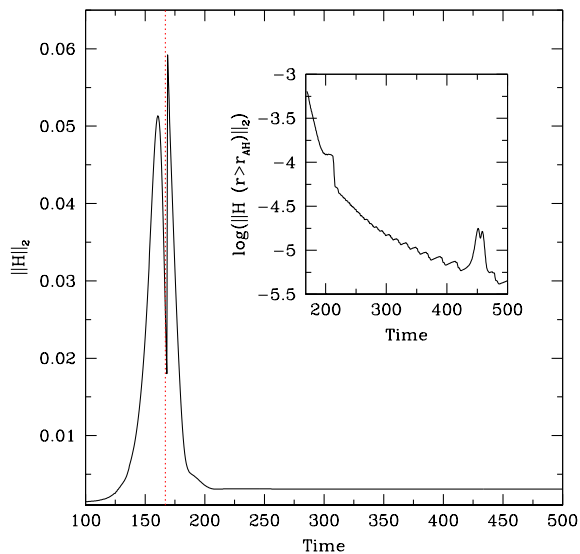} 
\caption{L2-norm of the Hamiltonian constraint in the collapse simulation. The 
vertical dashed line indicates the time of AH formation. The inset shows the 
L2-norm computed outside the AH.}
\label{fig8} 
\end{figure}

\subsection{Gravitational collapse of a marginally stable spherical 
relativistic star}
We next test the capability of the code to follow BH formation with the 
gravitational collapse to a BH of a marginally stable spherical relativistic 
star. For this test, we consider a $\kappa=100$, $N=1$ polytropic star with 
central rest-mass density $\rho_{c}=3.15\times 10^{-3}$, so that its 
gravitational mass is $M=1.64$ and its baryon rest-mass $M_{*}=1.79$. In order 
to induce the collapse of the star, we initially increase the rest-mass density
by 0.5$\%$. 

We present numerical results for a simulation of the gravitational collapse of 
a marginally stable spherical relativistic star performed with resolution of 
$\Delta r=0.125$. We use  the gauge conditions given in 
Eqs.~(\ref{1+log1})--(\ref{shift2}). We plot in Fig.~\ref{fig7} the time 
evolution of the normalized central density  until $t=300$ (upper panel), and of 
the mass of the AH in units of the ADM mass of the system until $t=500$ when we
stopped the simulation (lower panel). Overall, as the collapse proceeds the 
star increases its compactness, reflected in the increase of the central 
density as shown in the upper panel. The most unambiguous signature of the 
formation of a BH during the simulation is the formation of an AH. Once an AH 
is found by the AH finder, we monitor the evolution of the AH area, and also of
its mass which is plotted, in the lower panel of Fig.~\ref{fig7}. This panel 
shows that approximately at $t\sim 167$, an AH is first found and that the mass
of the AH relaxes to the ADM mass of the system. The difference in the ADM mass
and the mass of the AH at $t=500$, is about $0.2\%$.

In Fig.\ref{fig8} we plot the L2-norm of the Hamiltonian constraint in the 
collapse simulation. The vertical dashed line indicates the time of AH 
formation, and the inset shows the L2-norm computed outside the AH. The largest
violation of the constraint occurs during the AH formation, and afterwards the 
value of the L2-norm settles to $\sim 10^{-3}$. As in the case of a 
Schwarzschild BH, the L2-norm of the Hamiltonian constraint computed outside 
the AH is about two orders of magnitude smaller than the L2-norm computed in 
the whole grid. 

Results of this simulation indicate that the numerical scheme to integrate the 
evolution equations in time can handle accurately the transition between a 
regular spacetime (that of the star) and a irregular spacetime containing a 
puncture singularity at $r=0$.

\section{Conclusions}
\label{sec:conclusion}
In this paper we have presented a numerical code solving the BSSN equations in 
spherical symmetry and the general relativistic hydrodynamic equations written 
in flux-conservative form. A key feature of the code is that it uses a 
second-order PIRK method to integrate the evolution equations in time. This 
numerical scheme has proved to be crucial and sufficient to obtain the desired 
stability without the need for a regularization scheme at the origin. 

We have performed and discussed a number of tests to assess the accuracy and 
expected convergence of the code, namely a pure gauge wave, the evolution of a 
single BH, the evolution of spherical relativistic stars in equilibrium, and 
the gravitational collapse of a spherical relativistic star leading to the 
formation of a BH. We remark that, to our knowledge, we have presented the 
first successful numerical simulations of regular spacetimes (vacuum and 
non-vacuum) using the covariant BSSN formalism in spherical coordinates without
the need for a regularization algorithm at the origin (or without performing a 
spherical reduction of the equations~\cite{Garfinkle08,Bernuzzi10}).

In addition, curvilinear coordinate systems facilitate the use of non-uniform 
radial grids (i.e., logarithmic radial coordinate) to achieve the required high resolution near the origin 
while still keeping the outer boundaries sufficiently far away. This is 
particularly useful if one aims to study astrophysical phenomena like the 
gravitational collapse or the dynamics of accretion disks around BHs. Such 
approach is simpler, and likely computationally less expensive, than the 
adaptive mesh refinement techniques used in 3D codes in Cartesian coordinates.

We note that, unlike with the Fully Constrained Formulation~\cite{FCF} in which
some of the equations take an elliptic form and where a similar PIRK has been 
successfully tested, the BSSN formulation is purely hyperbolic, and yet the 
application of the PIRK method has proved very robust and provided the 
numerical stability necessary to perform long-term simulations of regular 
spacetimes in curvilinear coordinates. The work  we have presented also paves 
the way for future comparisons of the performance in curvilinear coordinates 
between the BSSN and the Fully Constrained Formulation system. Moreover, the 
application of the PIRK method to the BSSN equations in 3D in such coordinate 
systems should be rather straight forward, and we aim to investigate this in a 
future work. 

\section*{ACKNOWLEDGEMENTS}
We thank E. M{\"u}ller for his comments and careful reading of the manuscript.
P.M. acknowledges support by the Deutsche Forschungsgesellschaft (DFG) through 
its Transregional Centers SFB/TR 7 ``Gravitational Wave Astronomy''.
I. C.-C. acknowledges support from Alexander von Humboldt Foundation.

\appendix
\section{Detailed source terms included in the PIRK operators for the evolution
equations}
\label{appendix}
The evolution Eqs.~(\ref{a}), (\ref{b}), (\ref{K}), (\ref{Aa}), (\ref{Deltar}),
(\ref{X}), (\ref{1+log1})-(\ref{shift2}), are evolved using a second-order PIRK
method, described in Sec.~III. In this Appendix the source terms 
included in the explicit or partially implicit operators are detailed.

Firstly, the hydrodynamic conserved quantities, $a$, $b$, $X$, $\alpha$ and 
$\beta^r$, are evolved explicitly, i.e., all the source terms of the evolution 
equations of these variables are included in the $L_1$ operator of the 
second-order PIRK method.

Secondly, $A_a$ and $K$ are evolved partially implicitly, using updated values 
of $\alpha$, $a$ and $b$; more specifically, the corresponding $L_2$ and $L_3$ 
operators associated to the evolution equations for $A_a$ and $K$ are:
\begin{align}
	L_{2(A_a)} &= - \left(\nabla^{r}\nabla_{r}\alpha 
- \frac{1}{3}\nabla^{2}\alpha\right) 
+ \alpha\left(R^{r}_{r} - \frac{1}{3}R\right), \\
	L_{3(A_a)} &= \beta^{r}\partial_{r}A_{a} + \alpha K A_{a} 
- 16\pi\alpha(S_a - S_b), \\
	L_{2(K)} &= - \nabla^{2}\alpha, \\
	L_{3(K)} &= \beta^{r} \partial_{r}K  
+ \alpha(A_{a}^{2} + 2A_{b}^{2} + \frac{1}{3}K^{2}) \nonumber \\
& + 4\pi\alpha(E + S_{a} + 2S_{b}).
\end{align}

Then, $\hat{\Delta}^{r}$ is evolved partially implicitly, using updated values 
of $\alpha$, $a$, $b$, $\beta^r$, conformal factor, $A_a$ and $K$; more 
specifically, the corresponding $L_2$ and $L_3$ operators associated to the 
evolution equation for $\hat{\Delta}^{r}$ are:
\begin{align}
	L_{2(\hat{\Delta}^{r})} &= \frac{1}{a}\partial^{2}_{r}\beta^{r} 
+ \frac{2}{b}\partial_{r}\left(\frac{\beta^r}{r}\right)
+ \frac{\sigma}{3 a}\partial_{r}(\hat{\nabla}_m\beta^{m}) \nonumber \\
  & - \frac{2}{a}(A_{a}\partial_{r}\alpha + \alpha\partial_{r}A_{a}) 
- \frac{4\alpha}{r b}(A_{a}-A_{b}) \nonumber \\
	& + \frac{\xi \alpha}{a} \left[\partial_{r}A_{a} 
- \frac{2}{3}\partial_{r}K + 6A_{a}\partial_{r}\chi  \right. \nonumber \\
  & \left. + (A_{a}-A_{b})\left(\frac{2}{r} + \frac{\partial_{r}b}{b}\right)
\right], \\
	L_{3(\hat{\Delta}^{r})} &= \beta^{r}\partial_{r}\hat{\Delta}^{r}
- \hat{\Delta}^{r}\partial_{r}\beta^{r}
+ \frac{2\sigma}{3}\hat{\Delta}^{r}\hat{\nabla}_m\beta^{m} \nonumber \\
	& + 2\alpha A_{a}\hat{\Delta}^{r} - 8\pi j_{r} \frac{\xi \alpha}{a}.
\end{align}

Finally, $B^r$ is evolved partially implicitly, using updated values of 
$\hat{\Delta}^{r}$, i.e., 
$\displaystyle L_{2(B^r)} = \frac{3}{4}\partial_{t}\hat{\Delta}^{r}$ and 
$L_{3(B^r)} = 0$.


\bibliographystyle{apsrev}


\end{document}